\documentclass[journal=jacsat,manuscript=article]{achemso}

\usepackage{chemformula} 
\usepackage[T1]{fontenc} 
\usepackage{units}
\usepackage{hyperref}

\author{A. Pfeiffer}
\affiliation{ 
	Institut f\"ur Physik, Johannes Gutenberg-Universit\"at Mainz, 55099 Mainz, Germany
}

\affiliation{Graduate School of Excellence Materials Science in Mainz (MAINZ),
	Staudinger Weg 9, 55128 Mainz, Deutschland}

\author{R. M. Reeve}
\affiliation{ 
	Institut f\"ur Physik, Johannes Gutenberg-Universit\"at Mainz, 55099 Mainz, Germany
}
\affiliation{Graduate School of Excellence Materials Science in Mainz (MAINZ),
	Staudinger Weg 9, 55128 Mainz, Deutschland}

\author{K. Elphick}
\affiliation{Department of Electronic Engineering, University of York, Heslington, YO10 5DD York, UK}

\author{A. Hirohata}
\affiliation{Department of Electronic Engineering, University of York, Heslington, YO10 5DD York, UK}

\author{M. Kl\"aui}
\email{klaeui@uni-mainz.de}
\affiliation{ 
	Institut f\"ur Physik, Johannes Gutenberg-Universit\"at Mainz, 55099 Mainz, Germany
}
\affiliation{Graduate School of Excellence Materials Science in Mainz (MAINZ),
	Staudinger Weg 9, 55128 Mainz, Deutschland}

\title[]{Revealing the importance of interfaces for pure spin current transport}


\begin{document}


\begin{abstract}
Spin transport phenomena underpin an extensive range of spintronic effects. In particular spin transport across interfaces occurs in most device concepts, but is so far poorly understood. As interface properties strongly impact spin transport, one needs to characterize and correlate them to the fabrication method. Here we investigate pure spin current transport across interfaces and connect this with imaging of the interfaces. We study the detection of pure spin currents \textit{via} the inverse spin Hall effect in Pt and the related spin current absorption by Pt in Py-Cu-Pt lateral spin valves. Depending on the fabrication process, we either find a large (inverse) spin Hall effect signal and low spin absorption by Pt or vice versa. We explain these counter-intuitive results by the fabrication induced varying quality of the Cu/Pt interfaces, which is directly revealed \textit{via} a special scanning electron microscopy technique for interface imaging and correlated to the spin transport.

\end{abstract}

\maketitle

Spintronics aims to harness the electronic spin degree of freedom for a wide range of potential device functionalities that are underpinned by spin transport through interfaces such as magnetoresistance effects including giant magnetoresistance~\cite{Gruenberg1989,Baibich1988} and spin-torque manipulation of magnetic states.~\cite{Myers1999,Ralph2008,Manchon2019} While the structure and composition of the interfaces can be expected to strongly affect the spin transport,~\cite{Sanchez2013,Zhang2015,Dolui2017} a direct characterization of the systems is challenging since the relevant interfaces are often buried under other layers.  Hence they are not accessible with conventional surface sensitive imaging approaches and transmission electron microscopy does not allow for quantification of the quality of large areas. In particular the choice of the fabrication process can drastically modify the interfaces of the device, yet without such direct characterization of the interfaces the correlation between the fabrication method, interface quality and measured spin transport remains largely unknown. Recently, however, a technique has been pioneered which provides direct non-destructive access to buried interface characterization \textit{via} a scanning electron microscope (SEM).~\cite{Hirohata2016,Jackson2018} This technique opens up possibilities for such a direct connection between the spin transport and interfaces in relation to the fabrication recipes.

One particular prominent example where spin transport across interfaces plays a crucial role is the spin Hall effect (SHE), which provides the fascinating possibility to generate spin currents in nonmagnetic heavy metals (HM) with large spin orbit coupling.~\cite{Hirsch1999,Sinova2015} If an electric charge current is flowing in such a HM, a spin current is generated perpendicular to this charge current, with a polarization given by the vector product of the charge and spin current propagation direction.
The reciprocal inverse spin Hall effect (iSHE) describes the appearance of an electric charge current on spin current absorption such SHE materials.

In order to maximize spin transport across interfaces, in addition to the spin Hall angle $\theta_{\text{SH}}$ describing spin to charge conversion,~\cite{Sinova2015} other contributions including spin transparency~\cite{Zhang2015} and spin memory loss~\cite{Sanchez2013,Dolui2017,Berger2018} must be understood and optimized. In many studies such sources of spin relaxation are not considered, - yet they can significantly affect the measured signals.~\cite{Zhang2015,Berger2018} Spin transparency describes the relative transmission of different spin channels across the interface based on the spin mixing conductance.~\cite{Brataas2006} Spin memory loss describes the partial depolarization of the spin current caused by spin flip events as the spin current traverses the relevant interfaces.

Various techniques have been applied to determine $\theta_{\text{SH}}$ experimentally including spin pumping,~\cite{Ando2011,Zhang2013,Tao2018} spin Hall magnetoresistance~\cite{Nakayama2013} and spin orbit torque measurements.~\cite{Berger2018}
However, even in the case of Pt, the most widely studied spin Hall material, the determined spin Hall angles differ considerably between 0.01 and 0.20,~\cite{Wang2014,Zhang2015,Nguyen2016} depending on the particular study and the technique employed. These different techniques are based on different assumptions in particular concerning the nature of the relevant interfaces and hence their validity and applicability needs to be checked.

One particularly widely applied method to study spin transport through interfaces is the spin absorption method in a lateral spin valve device.~\cite{Niimi2012,Niimi2014,Sagasta2016,Laczkowski2017} In this geometry, two ferromagnetic electrodes (FM) are spatially separated but connected \textit{via} a nonmagnetic bridge (NM), which acts as a spin current conduit. The FM electrodes can be employed to inject and detect pure spin currents and if there is a further intermediate HM electrode present, additional spin current absorption by the HM can occur. By a comparison of the measured signals with and without a SHE intermediate electrode, as well as the generated iSHE voltage across the HM, models have been proposed to describe the spin transport and to determine the spin transport parameters of the materials, in particular the spin Hall angle  $\theta_{\text{SH}}$ and the spin diffusion length $\lambda_{\text{HM}}$ of the HM.~\cite{Niimi2012}

However, in these and similar devices,  the contribution of spin transparency and spin memory loss at the HM/NM and FM/NM interfaces and their connection to the interface structure is currently unclear. Of particular relevance is the role of the fabrication method in modifying spin transport \textit{via} the resulting possible changes to the interfaces, which however requires a direct comparison between the quality of the interfaces and the spin transport, what is so far missing.

In this work we investigate the spin transport across Cu/Pt and Cu/Py interfaces using the spin absorption method in a multi-terminal Pt-Py-Cu lateral spin valve. We fabricate two types of devices, for which different processes have been used to pattern the Cu bridge, leading to different interface properties and in particular different defect densities. For these two cases we compare the conventional non-local signal, the spin absorption and the (inverse) spin Hall effect signal for varying temperatures. Depending on the sample fabrication method, we either find a very large (i)SHE signal and nearly no spin absorption or a low (i)SHE signal but strong spin absorption. The results are discussed in terms of the properties of the interfaces of the devices. To reveal the origin, the interfaces are imaged \textit{via} a special scanning electron microscopy technique for buried interface characterization. Our findings demonstrate that the spin absorption method for determining spin transport parameters is not robust without further device characterization and highlight the sensitivity of spin transport to the interface properties that in turn are strongly governed by the fabrication method.

\begin{figure}
\includegraphics[width=0.75 \linewidth]{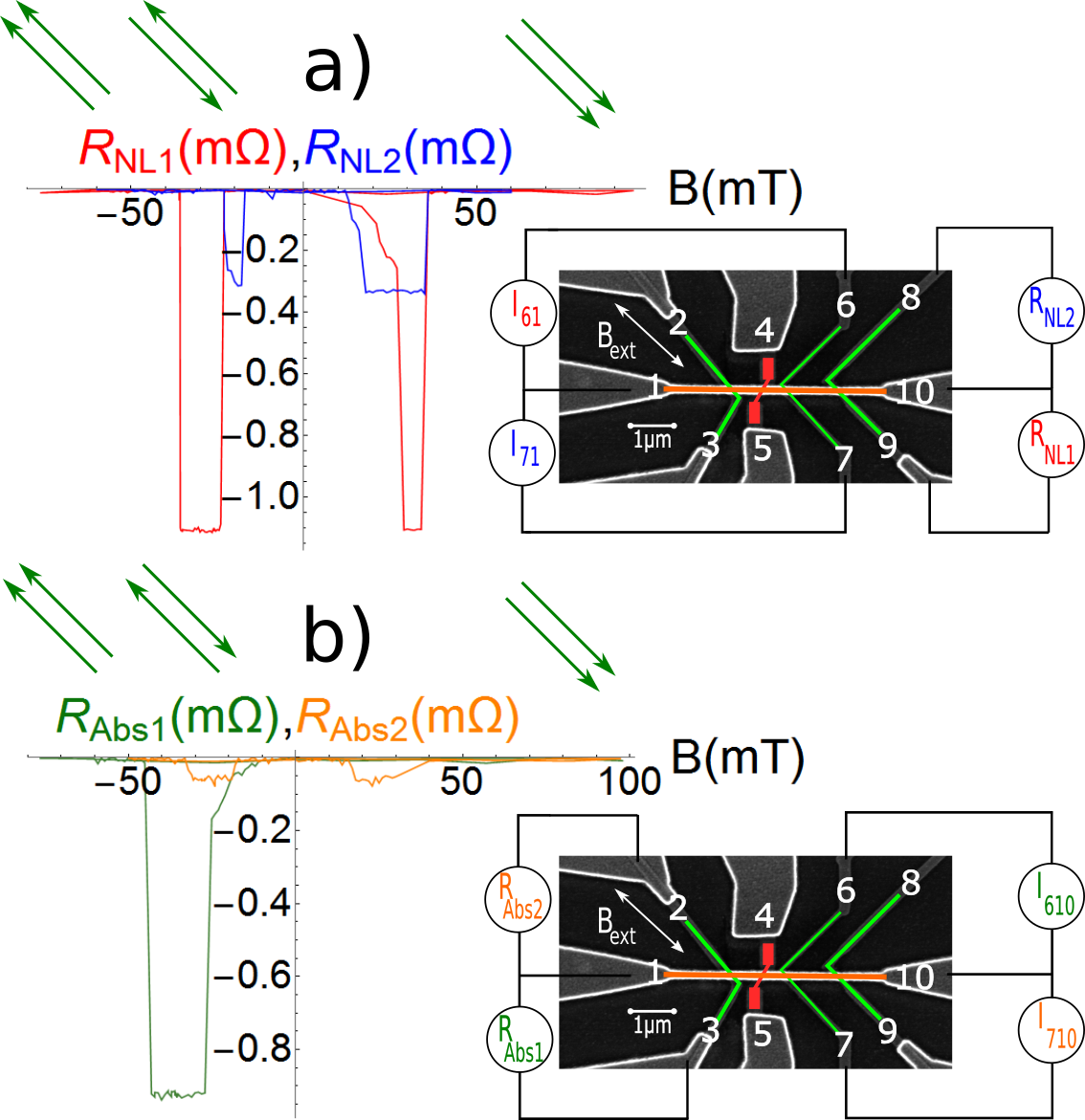}
\caption{Conventional non-local and spin absorption signal as a function of applied external field at \unit[4.2]{K} and the used probe configuration for the different recipes and signals.
	For both recipes, the sample consists of one Pt stripe (drawn in red), three Py wires (drawn in green) and one Cu bridge (drawn in orange). To generate both signals, the external field is swept along the easy axes of the Py wires and the magnetization orientations of the probed Py wires are drawn as green arrows above the plots. \textbf{a)} Conventional non-local signal of the sample fabricated by recipe 1 in red (\unit[1.10$\pm$0.01]{m$\Omega$}) and recipe 2 in blue  (\unit[0.34$\pm$0.01]{m$\Omega$}), together with the probe configuration. \textbf{b)} Spin absorption signal of the sample based on recipe 1 in green  (\unit[0.90$\pm$0.01]{m$\Omega$}) and recipe 2 in orange  (\unit[0.08$\pm$0.01]{m$\Omega$}), together with the probe configuration.}
\label{NLAbsbeide}
\end{figure}

Lateral spin valve samples with a kinked geometry as shown in the insets of \autoref{NLAbsbeide} are fabricated on a sapphire substrate by electron beam lithography (EBL) and lift-off techniques. In the first step, a \unit[100]{nm} wide and \unit[1]{$\mu$m} long  stripe is patterned, together with alignment markers and \unit[16]{nm} of Pt is deposited using magnetron sputtering (red stripe in \autoref{NLAbsbeide}). In the second step, three wires, one \unit[140]{nm} and two \unit[180]{nm} in width, are patterned perpendicularly to the Pt stripe and \unit[25]{nm} of Py is deposited by ultrahigh vacuum (UHV) thermal evaporation (green wires in \autoref{NLAbsbeide}). After the deposition and the lift-off processing of the Py wires, the substrate has been cut in two \unit[5$\cdot$10]{mm$^2$} pieces. For the patterning of the nonmagnetic bridge, two different recipes have been used.
\begin{itemize}
	\item \textbf{Recipe 1:}  \unit[300]{nm} of poly(methyl methacrylate) (PMMA) 950K A4 has been spun onto the chip and the EBL has been performed using \unit[20]{kV} acceleration voltage of the primary beam.
	\item \textbf{Recipe 2:} Firstly \unit[150]{nm} of methyl  methacrylate (MMA) EL6 has been spun and as a second resist, \unit[300]{nm} of PMMA 950 A4 has been used. The EBL has been performed using \unit[10]{kV} acceleration voltage of the primary beam.
\end{itemize}
For both recipes, the baking time of the resist(s) is \unit[90]{seconds} at \unit[180]{$^\circ$C} on a hotplate. The development has been performed for both recipes using one part of methyl isobutyl ketone (MIBK) diluted in three parts of isopropyl alcohol (IPA) for \unit[45]{seconds}. The used exposure doses for the two recipes have been independently optimized in order to yield low ohmic electric interface resistances in the \unit[m$\Omega$]-range.
In the last step, \textit{in-situ} argon milling is used to clean the interfaces of the Py wires and the Pt stripe for both recipes at the same time with the substrates mounted on the sample holder next to each other and with the same orientation with respect to the argon gun. Finally, a \unit[170]{nm} wide (for recipe 1) and a \unit[190]{nm} wide (for recipe 2) and \unit[85]{nm} thick Cu bridge has  been deposited \textit{via} UHV thermal evaporation, together with electric contacts (orange wire in \autoref{NLAbsbeide}).

To measure the different spin transport signals, namely the conventional non-local, the spin absorption and the (inverse) spin Hall effect signal at a temperature of \unit[4.2]{K}, an alternating current of \unit[1.0]{mA}  with a frequency of \unit[2221]{Hz} is applied. To generate this current, an alternating voltage of \unit[5]{V} amplitude has been applied and a \unit[5000]{$\Omega$} pre-resistor has been used before the sample to act as a current source. Since our nanowires have low resistances (some \unit[hundred]{$\Omega$}), we can assume the same current for all temperatures and neglect the small variations of the sample resistance which are on the order of \unit[200]{$\Omega$} for the central Py and less than \unit[40]{$\Omega$} for the left and right Py wire and the Pt stripe.

For the studied temperature dependence of the signals based on recipe 2, a pre-resistor of \unit[1100]{$\Omega$} has been used in order to apply higher currents. However, since the probe configuration has been changed, the left and right Py wire as well as the Pt stripe act as injector electrodes which have resistances below \unit[120]{$\Omega$} at room temperature and less than \unit[80]{$\Omega$} at \unit[4.2]{K}. Therefore the same current is assumed for all temperatures and  small variations of the sample resistance with varying temperature are neglected.

To measure the conventional non-local signal for samples fabricated by recipe 1 (2), we apply the current between contact 6 (7) as the top (bottom) part of the central Py wire and contact 1 as the left end of the Cu bridge. The non-local voltage is measured between contact 9 (8) as the bottom (top) end of the right Py wire and contact 10 as the right end of the Cu bridge.
To measure the spin absorption signal, the current is applied between contact 6 (7) and contact 10 and the non-local voltage is measured between contact 3 (2) as the bottom (top) part of the left Py wire and contact 1.

To generate both the conventional non-local and the spin absorption signal, an external field is swept between \unit[$-$100]{mT} and \unit[$+$100]{mT} parallel to the easy axes of the Py wires, as indicated in \autoref{NLAbsbeide}. 
The non-local resistance $R_{\text{NL}}$ and the (inverse) spin Hall effect resistance $R_{\text{(i)SHE}}$ are defined as the measured voltages, divided by the applied current. 
The error bars for the different temperature dependent curves are calculated as $ \Delta_{\text{tot}}=\sqrt{(\Delta_{\text{AP}})^2+( \Delta_\text{P})^2}$, with $\Delta_{\text{AP,P}}$ as the standard derivation of the signals for high and low spin signal states. As also mentioned in the main text, the reduction of the spin signals \textit{via} spin absorption is calculated as (1-(R$_{\text{Abs}}$/R$_{\text{NL}}$)).

Firstly, we compare the conventional non-local spin signals with spin current injection and detection in Py electrodes for the samples based on the two different recipes shown in \autoref{NLAbsbeide}a). We observe a signal of \unit[1.10$\pm$0.01]{m$\Omega$} for the sample based on recipe 1 (red curve) and a signal of \unit[0.34$\pm$0.01]{m$\Omega$} (blue curve) for the sample fabricated by recipe 2. We find that samples fabricated by recipe 1 consistently yield approximately a factor 3 higher spin signals than samples based on recipe 2. Since the electric charge current interface resistances are very similar for samples based on the different recipes (in the \unit{m$\Omega$}-range),  these differences in the signals cannot be explained by different charge transport interface resistances. Additionally we emphasize that for all measurements, the different possible injector/detectors permutations have been checked, with a maximum variation of the different signals for the different configurations within one device of \unit[25]{\%}. In this work we always show the highest measured spin signals, which accounts for the different indicated probe configurations for the different samples based on the different recipes.

Next we compare the spin absorption strength, which is a measurement of the conventional non-local spin signal after the (partial) spin current absorption by Pt, for the samples based on the two recipes shown in \autoref{NLAbsbeide}b). Here we find even larger differences. For the sample based on recipe 1, the spin  absorption signal is \unit[0.90$\pm$0.01]{m$\Omega$} (green curve) and thus about \unit[20]{\%} smaller than the conventional non-local signal. Hence within our variations of \unit[25]{\%} for the injector/detector configurations, we do not determine significant spin absorption at the Pt/Cu interface for the sample fabricated by recipe 1. For the sample based on recipe 2 however, we observe a spin absorption signal of \unit[0.08$\pm$0.01]{m$\Omega$} (orange curve), resulting in a reduction of \unit[76$\pm$3]{\%} (calculated as (1-(R$_{\text{Abs}}$/R$_{\text{NL}}$))) of the non-local signal from the case without the intermediate Pt electrode and therefore a large expected spin absorption by the Pt electrode. The determined results for the sample based on recipe 2 with a reduction of the signal of \unit[76$\pm$3]{\%} for Pt as an absorber material with large spin orbit coupling agrees well with findings in the literature, where the absorption of various SHE materials including Pt~\cite{Sagasta2016,Pham2016}, CuBi~\cite{Niimi2014} and AuTa~\cite{Laczkowski2017} has been studied using this approach in lateral spin valves. However, what is surprising in our study is the lack of differences in the conventional non-local and the spin absorption signal for the sample based on recipe 1, which calls for further investigation. In order to understand  the differences we next compare the spin absorption signals to the (inverse) spin Hall effect signals in the heavy metals.

To measure the inverse spin Hall effect signal, the same applied current has been used and the field is swept as before along the easy axes of the Py wires. Since the generated charge current in the Pt stripe due to the iSHE is
\begin{align}
	J_\text{{iSHE}} \propto J_s \times \sigma,
\end{align}	
the polarity of the generated charge current $J_\text{{iSHE}}$ changes sign on changing the orientation of the spin current $J_s$. In order to switch $J_s$, the injector magnetization is reversed by sweeping the external magnetic field.

As in our previously used geometry,~\cite{Pfeiffer2018} we have patterned the Pt stripe and the Py wire perpendicularly to each other which provides the maximum changes in the spin signal \textit{via} sweeps of the external field parallel to the easy axis of the FM wire. As a result, the two non-local resistance levels corresponding to two magnetic states are stable at remanence. Furthermore, the required fields to fully saturate the magnetization are much lower compared to previous publications where the heavy metal and the magnetic wires were often all oriented parallel to each other.~\cite{Niimi2014,Sagasta2016}

\begin{figure}
\includegraphics[width=0.99\linewidth]{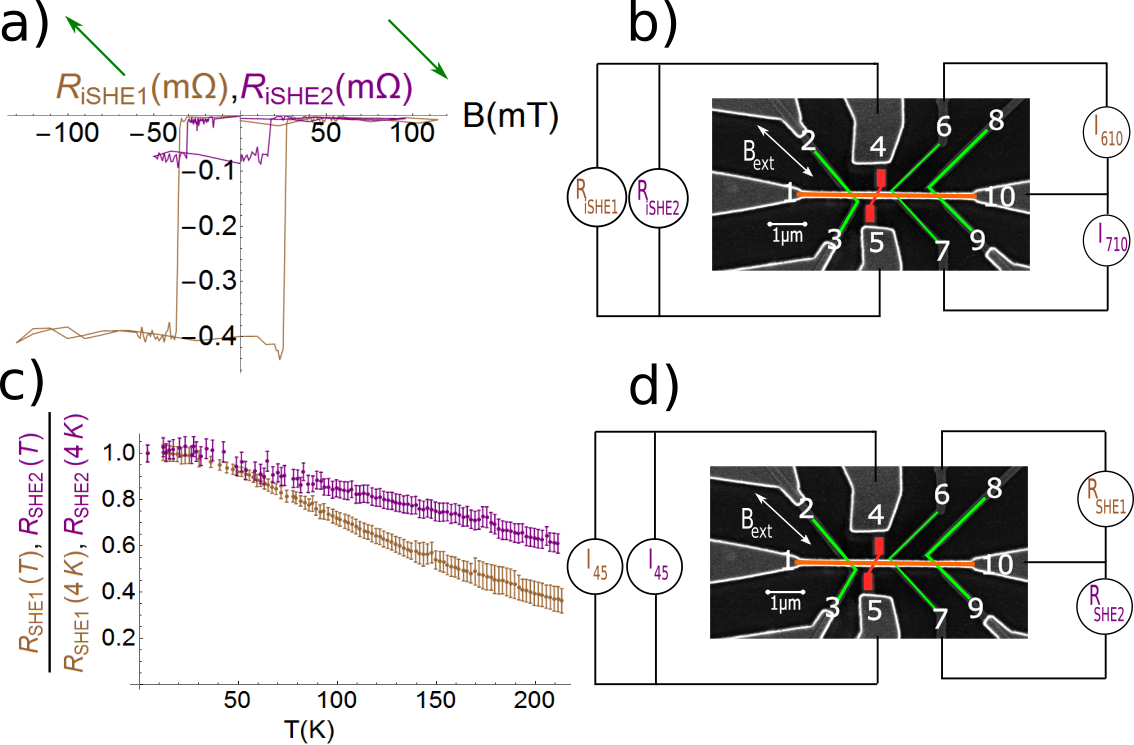}
\caption{
	\textbf{a)} Inverse spin Hall effect signal as a function of the applied external field. We find for both recipes two distinct states which depend on the magnetization orientation of the injector, indicated by the green arrows. For the sample based on recipe 1 a signal of \unit[0.40$\pm$0.01]{m$\Omega$} (brown curve) and for the sample based on recipe 2 a signal of \unit[0.08$\pm$0.01]{m$\Omega$} (purple curve) is determined. \textbf{b)} Probe configuration for the studied inverse spin Hall effect signals.
	\textbf{c)} Normalized  spin Hall effect signal of the sample fabricated by recipe 1 (brown) curve and of the sample fabricated by recipe 2 (purple curve) in the temperature range between \unit[4.2]{K} and \unit[200]{K}. For temperatures between \unit[50]{K} and \unit[200]{K}, a larger decrease of the normalized signal for the sample fabricated by recipe 1 is observed. \textbf{d)} Probe configuration for the studied spin Hall effect signals.}
\label{ISHE4KTemp}
\end{figure}

As presented in \autoref{ISHE4KTemp}a), we find for the sample fabricated by recipe 1 an inverse spin Hall effect signal of \unit[0.40$\pm$0.01]{m$\Omega$} (brown curve) and for the sample based on recipe 2 a signal of \unit[0.08$\pm$0.01]{m$\Omega$} (purple curve). The size of the signal for the sample based on recipe 2 is consistent with the measured spin signal in our previous work,~\cite{Pfeiffer2018} where also recipe 2 has been used for the Cu bridge processing. 
This difference of the iSHE signals as compared with the difference of the spin absorption signals, however, is counter-intuitive:

One would expect that a large reduction of the spin signal by spin absorption into the Pt stripe should be connected with a large (inverse) spin Hall effect signal in the Pt, if both signals are based on the same spin current. 
From these results, we conclude that the size of the so-called spin absorption signal is not only related to the intrinsic properties of the Pt electrode. Rather the Pt/Cu interface properties are of key importance and additional contributions need to be taken into account which reduce the spin current without contributing to the inverse spin Hall effect, such as interface spin memory loss.~\cite{Sanchez2013,Dolui2017,Berger2018,Tao2018}

To check this, we probe the normalized temperature dependence of  the spin Hall effect for samples based on the two recipes, as  shown in \autoref{ISHE4KTemp}c)). Due to Onsager reciprocity,~\cite{Kimura2007,Jacquod2012} varying the probe configuration does not change the signal, as confirmed in our measurements.~\cite{Pfeiffer2018} We find for temperatures between \unit[50]{K} and \unit[200]{K} a stronger decrease of the normalized spin Hall effect signal measured for the sample based on recipe 1 compared to recipe 2. These differences can be explained by a strong temperature independent reduction of the spin current, which partially masks the temperature dependent contribution.

As a result, the decrease of the spin diffusion length in the Cu bridge with increasing temperature~\cite{Kimura2008} as expected from Elliot-Yafet theory,~\cite{Elliot1954,Yafet1963} which is usually sufficient to explain the temperature behaviour of the inverse spin Hall effect signal in this temperature range, is less dominant for samples fabricated by recipe 2 compared to samples fabricated by recipe 1. If the Pt/Cu interfaces for the two recipes are indeed significantly different (despite the very similar electric interface resistances), we expect that these differences should also clearly affect the temperature behaviour of both the conventional non-local and the spin absorption signal for samples based on the different recipes. This needs to be checked next.

\begin{figure}
\includegraphics[width=0.9 \linewidth]{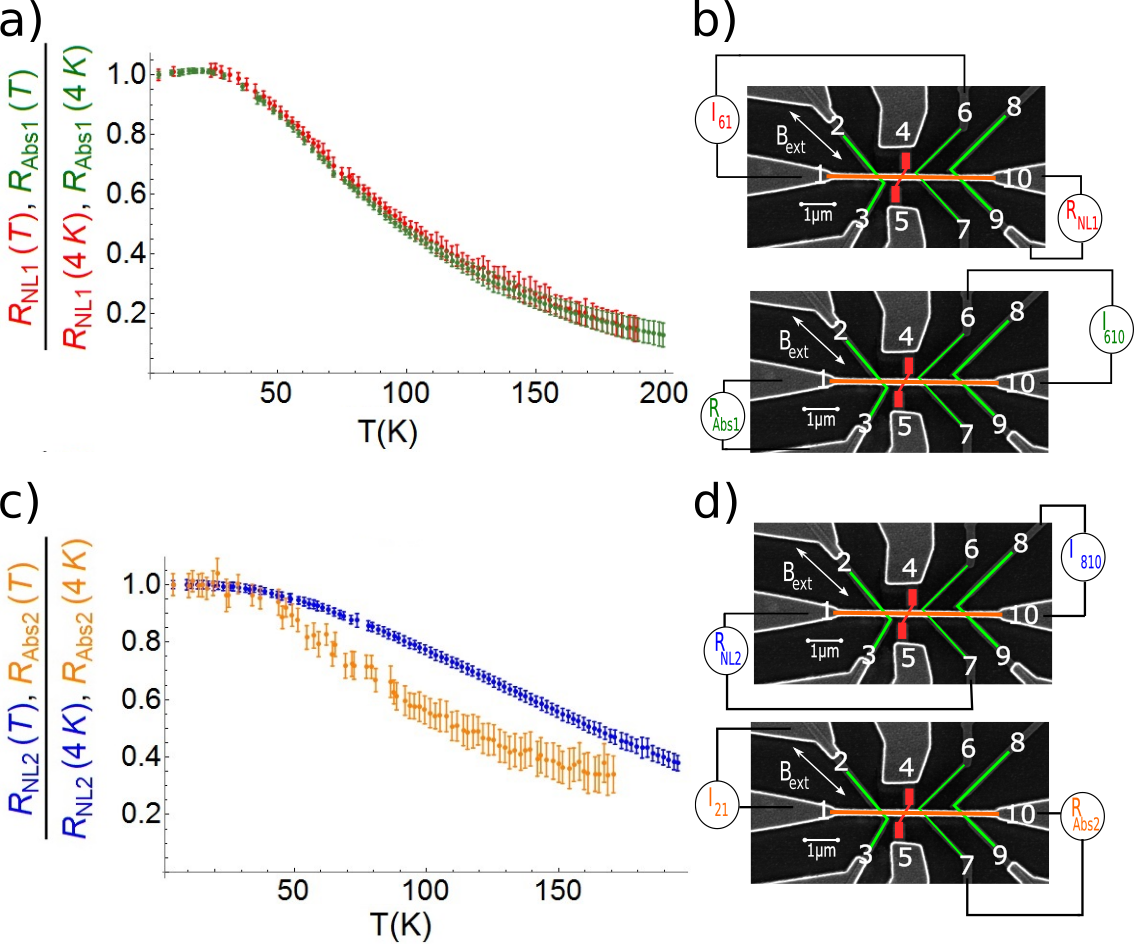}
\caption{Temperature dependence of the normalized conventional non-local signal and the spin absorption signal of the samples based on the different recipes. 
	\textbf{a)} Temperature dependence of the conventional non-local signal (red curve) and the spin absorption signal (green curve) for the sample based on recipe 1. Within the error bars, the two curves are equivalent. 
	\textbf{b)} Probe configuration of the two studied signals for the sample based on recipe 1.
	\textbf{c)} Temperature dependence of the conventional non-local signal (blue curve) and the spin absorption signal (orange curve) for the sample fabricated by recipe 2. In the temperature range between \unit[50]{K} and \unit[200]{K} we find a stronger decrease of the spin absorption signal with increasing temperature. 
	\textbf{d)} Probe configuration of the two studied signals for the sample based on recipe 2.  }
\label{NLAbsTemp}
\end{figure}

We now compare the temperature dependencies for the normalized conventional non-local and the spin absorption signal for both recipes, as plotted in \autoref{NLAbsTemp}. As expected from the comparison of the two signals at \unit[4.2]{K}, the signals of the sample based on recipe 1 are equivalent within the error bars. This is different for the sample based on recipe 2, where a much stronger reduction of the spin absorption signal with increasing temperature is found. In previous publications, differences in the temperature behaviour during the spin transport have been usually attributed to effects such as increased surface scattering~\cite{Kimura2008,Villamor2013} or the Kondo effect.~\cite{OBrien2014,Batley2015,Kim2017} Since here the Cu conduit and the Py/Cu interfaces are the same for a given recipe, all changes between the non-local signal and the spin absorption signal must be connected to the additional Pt/Cu interface. 

These large differences in the temperature dependence for the two signals for the sample fabricated by recipe 2, combined with the identical temperature dependence of the signals of the sample based on recipe 1,  support our previous findings concerning the differences in the (inverse) spin Hall effect signals. A significant amount of the generated spin current in the sample based on recipe 2 is lost due to spin flip events at the interface and does not contribute to the (i)SHE signal.

To directly probe the structural nature of the different relevant Py/Cu and Pt/Cu interfaces for devices made \textit{via} the two recipes directly and thus to reveal the origin of the different results of the spin transport measurements, we additionally characterize the relevant interfaces \textit{via} a special SEM technique. This technique allows for non-destructive buried interface imaging by employing a decelerated electron beam, as explained in detail in our previous work.~\cite{Hirohata2016,Jackson2018} 

Individual SEM images have been taken using the commercial SEM ``JEOL JSM 7800F''. These images have been taken using different acceleration voltages, which are selected based on the results of a ``CASINO'' electron trajectory simulation.~\cite{Drouin2007} The selective voltages for analysing the Cu/Py interface are \unit[4.8]{kV} and \unit[5.1]{kV}, while for analysing the Cu/Pt interface,  voltages of \unit[4.9]{kV} and \unit[5.1]{kV} have been used. An upper electron detector (UED) has been used in order to maximize the back-scattered electron signal. The images, taken with different acceleration voltages have been subsequently compared using a ``MATLAB'' script. In this manner, the contrast and alignment of the images has been re-adjusted and a processed image is generated.

\begin{figure*}
\includegraphics[width=0.50 \linewidth]{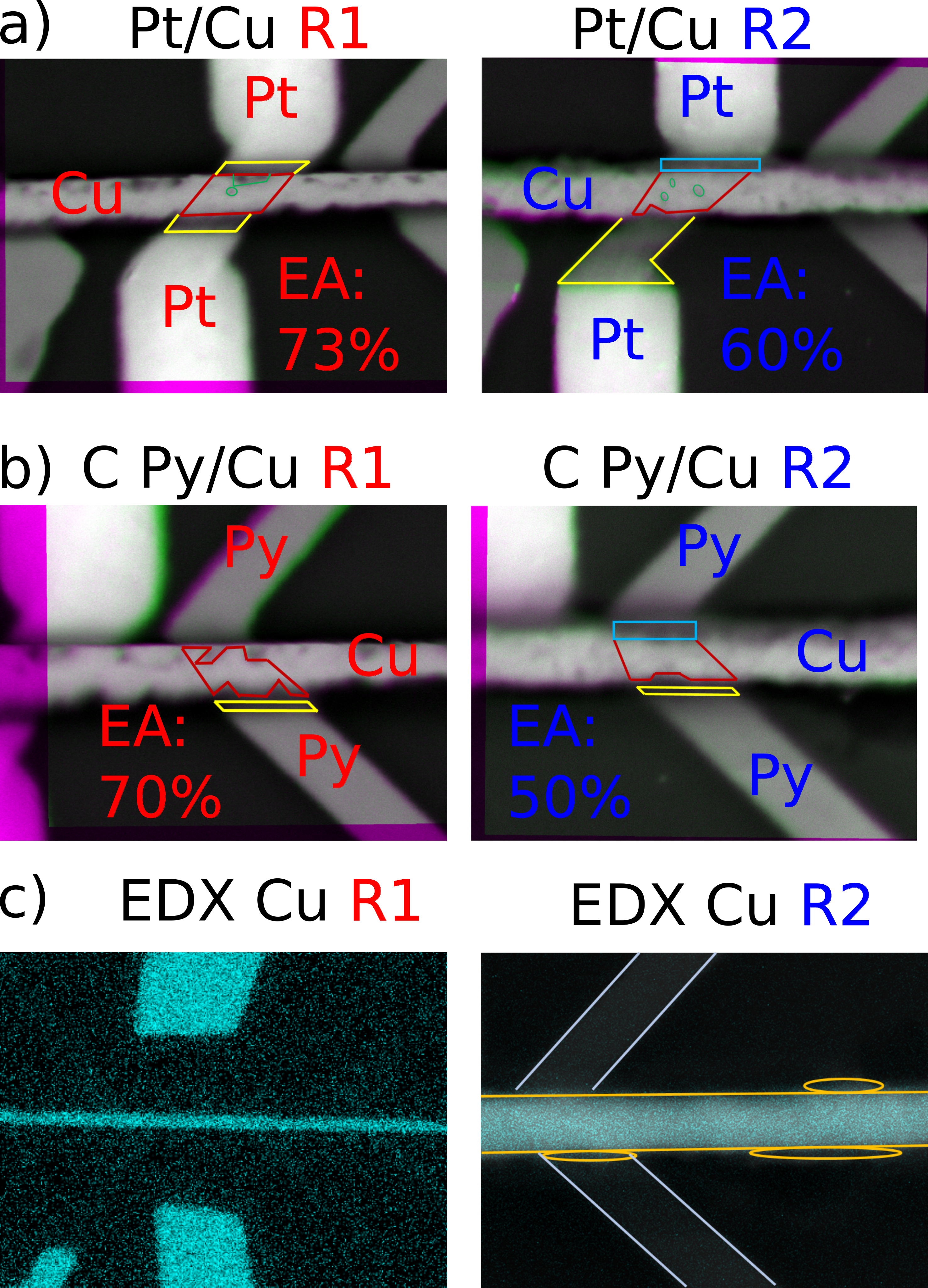}
\caption{ \textbf{a), b)} Imaging of buried Pt/Cu and Py/Cu interfaces of samples based on the two different recipes. Regions marked by the red boxes indicate an effective area with a low amount of defects. Single defects within the red area are marked by green circles. We find for samples based on recipe 2 shadow regions (marked by the blue boxes) at the top edge of the Cu/Pt and the Cu/Py interfaces, reducing significantly the effective defect free interface area EA. The yellow marked areas indicate the variation of the thickness of the Pt stripe and the Py wires, caused by the in-situ milling procedure.  
	\textbf{c)} Performed EDX measurements for the Cu content of the samples based on the different recipes. We find for samples based on recipe 2 additional Cu content at the edges of the Cu wire (marked by the yellow ellipses), consistent with the lower quality of the interfaces of samples based on recipe 2 compared to samples fabricated by recipe 1. \\
}
\label{VergleichInterfaces}
\end{figure*}

We present here representative images of the Pt/Cu and the central Py/Cu interfaces, since the left and the right Py/Cu interface show analogous results to the central one.
As shown in \autoref{VergleichInterfaces}, we observe significantly less inhomogeneities for samples fabricated by recipe 1 compared to samples based on recipe 2, which can be quantified \textit{via} an ``effective defect free interface area'' EA. The effective defect free interface area can be understood as the area (marked by red boxes) without detectable inhomogeneities, divided by the total area.
We emphasize that these effective areas should not be mistaken with the contact area of the interfaces, since the non-local signal scales reciprocally with the contact area,~\cite{Kimura2006} which is not the case for the EA described here. Rather the EA corresponds to interface regions where effective spin transport is expected.
Within a clean area, we mark single defects as green circles. The areas marked in yellow are related to the varying thickness of the Pt and Py wires, caused by the in-situ milling procedure which decreases the thickness of the previously exposed parts of the Pt stripe and the Py wires. 

We find for samples based on recipe 1 much more homogeneous Pt/Cu and Py/Cu interfaces, resulting in a \unit[73]{\%} EA for the Pt/Cu and a \unit[70]{\%} EA for the central Py/Cu interface.
In particular for samples based on recipe 2 we observe at the top Pt/Cu and the top Py/Cu edge a shadow region (marked as a blue box in the plots), which is also found at the left and right Py/Cu interfaces. These shadow regions significantly reduce the determined effective area down to \unit[60]{\%} for the Pt/Cu and \unit[50]{\%} for the central Py/Cu interface, reducing the overall interface quality for samples fabricated by recipe 2. Our results lend themselves to the explanation that significantly more spin relaxation at the interfaces is generated for samples based on recipe 2. Consequently both a lower (i)SHE signal and a larger reduction in the non-local signal is found, compared to samples based on recipe 1 with better interfaces.

The worse interface quality for samples based on recipe 2 compared to samples based on recipe 1 is supported by energy-dispersive X-ray spectroscopy (EDX) measurements, as presented in \autoref{VergleichInterfaces}c).
While for samples fabricated by recipe 1, the EDX results are as expected based on the sample design, we find for samples based on recipe 2 some additional Cu content at the Pt/Cu and Py/Cu edges, which is marked with yellow ellipses in the plot. From these measurements we conclude that although the lift-off based on a  double layer resist (recipe 2) is significantly easier compared to the lift-off based on a single layer resist (recipe 1), recipe 2 leads to Cu content at undesired positions, which is consistent with the lower quality of the interfaces of samples based on recipe 2 compared to samples fabricated by recipe 1.

As mentioned, double-layer resists are usually employed to facilitate the lift-off process for thick films. However, our work demonstrates that the device performance depends significantly on the used patterning process and deposition conditions since they lead to different interface qualities, which are not apparent from conventional characterization. This has direct consequences for the determination of key transport parameters such as $\theta_{\text{SH}}$ and $\lambda_{\text{HM}}$, describing the HM influence on the spin absorption and the (inverse) spin Hall effect signal. We  determine for samples based on recipe 2 the spin diffusion length and the spin Hall angle of Pt to be \unit[3]{nm} and \unit[0.16]{\%}, respectively at \unit[4.2]{K}. 
For samples fabricated by recipe 1 (assuming R$_{\text{Abs1}}$/R$_{\text{NL1}}$=0.8), we evaluate $\lambda_{\text{HM}}$ to be \unit[19]{nm} while   $\theta_{\text{SH}}$ is determined to be \unit[1.6]{\%} at \unit[4.2]{K} by using the same method as Sagasta \textit{et al}.~\cite{Sagasta2016} Hence depending on the interface quality, there is an order of magnitude difference in the determined effective parameters, similar to the spread seen in the reports from different groups in the literature.~\cite{Wang2014,Zhang2015,Nguyen2016}

To explain the strong reduction of the spin absorption signal together with the low (i)SHE signal measured in samples based on recipe 2, different possible contributions are considered.  One possibility is that residual resist is present at the interface and the contamination and associated disorder in those regions lead to enhanced spin flip scattering, leading to spin memory loss. Furthermore the structural disorder could modify the bond-modelling and related spin transparency of the interface.

The situation is different for samples based on recipe 1, where we find that the interfaces are of significantly higher quality. Additionally in these samples, no differences of the temperature dependence between conventional non-local and spin absorption signal have been seen. These results, which reveal strong differences depending on the interface homogeneity, suggest that spin transparency and spin memory loss are crucial and must be taken into account to explain the surprising high (i)SHE signal and the surprising low spin absorption.
Our results for these samples suggest that there is little spin memory loss at the Pt/Cu interface for samples based on recipe 1 and therefore no significant reduction of the spin absorption signal compared to the conventional non-local signal, in contrast to samples based on recipe 2. However, it is important to note that the absence of spin memory loss does not necessarily directly lead to a large (i)SHE signal, depending on the transparency of the interface. Due to the finite spin diffusion length in Pt, only the spin accumulation that has passed across the Pt/Cu interface can be scattered and consequently contribute to the (inverse) spin Hall effect.

Based on the strong variations of the determined effective spin transport parameters of the different systems, we conclude that the spin absorption method faces challenges for robust determination of the spin Hall properties due to its strong sensitivity to the different interfaces. Experimental methods which are based on thin films and do not require a multi-step lift-off process, \textit{e.g.} spin pumping~\cite{Ando2011,Zhang2013,Tao2018}, spin torque ferromagnetic resonance~\cite{Zhang2015} or spin orbit torque measurements~\cite{Berger2018} might be more suitable for an accurate determination of $\theta_{\text{SH}}$ and $\lambda_{\text{HM}}$. While also in these measurements contributions such as spin transparency and spin memory loss need to be considered, possible problems with residual resist and other fabrication related inhomogeneities at the interfaces are less relevant.

We furthermore conclude that in our samples in addition to the intrinsic properties of Pt ($\lambda_{\text{HM}}$ and $\theta_{\text{SH}}$), which are relevant to explain the device behaviour, also the influence of the relevant interfaces and especially contributions such as spin transparency and spin memory loss play a decisive role. In particular characterization methods that can directly reveal the loss of spin information at the interfaces are crucial for a proper determination of spin absorption and the (inverse) spin Hall effect. For a proper characterization of the interfaces, both temperature dependent spin transport and buried interface imaging are necessary since the subtle differences that lead to different spin transport may not be revealed by electrical characterization. Especially the correlation between the fabrication method and the resulting interfaces characterized by buried interface imaging and spin transport measurements provides invaluable insights into the device performance and reveals the strong sensitivity of spin transport to the fabrication recipe. In order to maximize the spin signals, next to a careful tailoring of the relevant interfaces, material combinations with minimized spin memory loss and maximized spin transparency, \textit{e.g.} by suitable band structure matching,~\cite{Lidig2019} are promising.

In summary, we have studied multi-terminal Pt-Py-Cu based lateral spin valves, which allow us to compare the spin absorption signal with the size of the (inverse) spin Hall effect for two fabrication recipes for the Cu conduit patterning. Very similar charge transport properties of the interfaces of the samples for the two recipes are found. However, we observe drastically different conventional non-local, spin absorption and (inverse) spin Hall effect signals for samples based on the different recipes. For samples fabricated by the first recipe, where a single PMMA-resist layer has been used for processing, a very large (inverse) spin Hall effect signal is found but no significant spin absorption at the Pt/Cu interface is observed. For samples based on the second recipe, where a dual MMA-PMMA-resist layer has been used, we observe a low (inverse) spin Hall effect signal but find a reduction of the non-local signal of \unit[76$\pm$3]{\%} at \unit[4.2]{K}.

These large differences of the signals for the different fabrication recipes are explained by interface spin loss and spin relaxation at the Pt/Cu interface due to the different qualities of the interfaces.
These effects crucially changes the spin properties but do not affect the electrical properties of our devices. By performing direct imaging of the buried Pt/Cu and Py/Cu interfaces, we observe significantly higher quality interfaces for samples based on recipe 1 compared to recipe 2. Thus we are able to directly link the obtained spin signal with the imaged interface quality, which in-turn is determined by the fabrication method.

Our results clearly indicate that for a full understanding of spin transport through interfaces, not only the electric charge transport but additionally the interface spin transport properties are crucial and need to be tailored carefully.

One prominent consequence in this work is the fact that we determine strongly varying effective spin transport parameters of identically deposited Pt ($\lambda_{\text{HM}}$ of \unit[19]{nm} for recipe 1 and \unit[3]{nm} for recipe 2 and $\theta_{\text{SH}}$ of \unit[1.6]{\%} for recipe 1 and \unit[0.16]{\%} for recipe 2) for the two fabrication recipes. We conclude that the widely used spin absorption method is not always robust due to its strong sensitivity to the interface quality, which is not revealed in conventional electrical sample characterization or conventional imaging.
Overall, the work demonstrates the vital role of interface structure in spin transport by both correlating spin transport properties and the interface structural quality obtained from buried interface imaging and we find that the interface quality is strongly dependent on the fabrication techniques.

\begin{acknowledgement}

We acknowledge financial support of the SFB/TRR 173 Spin+X: spin in its collective environment funded by the Deutsche Forschungsgemeinschaft (DFG, German Research Foundation) Project No. 290396061/TRR173, as well as the Graduate School of Excellence Materials Science in Mainz (No. GSC266). \\
This work was partially supported by JST CREST (No. JPMJCR17J5).

\end{acknowledgement}

\providecommand{\latin}[1]{#1}
\makeatletter
\providecommand{\doi}
{\begingroup\let\do\@makeother\dospecials
	\catcode`\{=1 \catcode`\}=2 \doi@aux}
\providecommand{\doi@aux}[1]{\endgroup\texttt{#1}}
\makeatother
\providecommand*\mcitethebibliography{\thebibliography}
\csname @ifundefined\endcsname{endmcitethebibliography}
{\let\endmcitethebibliography\endthebibliography}{}


\begin{mcitethebibliography}{39}
	\providecommand*\natexlab[1]{#1}
	\providecommand*\mciteSetBstSublistMode[1]{}
	\providecommand*\mciteSetBstMaxWidthForm[2]{}
	\providecommand*\mciteBstWouldAddEndPuncttrue
	{\def\EndOfBibitem{\unskip.}}
	\providecommand*\mciteBstWouldAddEndPunctfalse
	{\let\EndOfBibitem\relax}
	\providecommand*\mciteSetBstMidEndSepPunct[3]{}
	\providecommand*\mciteSetBstSublistLabelBeginEnd[3]{}
	\providecommand*\EndOfBibitem{}
	\mciteSetBstSublistMode{f}
	\mciteSetBstMaxWidthForm{subitem}{(\alph{mcitesubitemcount})}
	\mciteSetBstSublistLabelBeginEnd
	{\mcitemaxwidthsubitemform\space}
	{\relax}
	{\relax}
	
	\bibitem[Binasch \latin{et~al.}(1989)Binasch, Gr\"unberg, Saurenbach, and
	Zinn]{Gruenberg1989}
	Binasch,~G.; Gr\"unberg,~P.; Saurenbach,~F.; Zinn,~W. Enhanced
	magnetoresistance in layered magnetic structures with antiferromagnetic
	interlayer exchange. \emph{Phys. Rev. B} \textbf{1989}, \emph{39},
	4828--4830\relax
	\mciteBstWouldAddEndPuncttrue
	\mciteSetBstMidEndSepPunct{\mcitedefaultmidpunct}
	{\mcitedefaultendpunct}{\mcitedefaultseppunct}\relax
	\EndOfBibitem
	\bibitem[Baibich \latin{et~al.}(1988)Baibich, Broto, Fert, Van~Dau, Petroff,
	Etienne, Creuzet, Friederich, and Chazelas]{Baibich1988}
	Baibich,~M.~N.; Broto,~J.~M.; Fert,~A.; Van~Dau,~F.~N.; Petroff,~F.;
	Etienne,~P.; Creuzet,~G.; Friederich,~A.; Chazelas,~J. Giant
	Magnetoresistance of (001){F}e/(001){C}r Magnetic Superlattices. \emph{Phys.
		Rev. Lett.} \textbf{1988}, \emph{61}, 2472--2475\relax
	\mciteBstWouldAddEndPuncttrue
	\mciteSetBstMidEndSepPunct{\mcitedefaultmidpunct}
	{\mcitedefaultendpunct}{\mcitedefaultseppunct}\relax
	\EndOfBibitem
	\bibitem[Myers \latin{et~al.}(1999)Myers, Ralph, Katine, Louie, and
	Buhrman]{Myers1999}
	Myers,~E.~B.; Ralph,~D.~C.; Katine,~J.~A.; Louie,~R.~N.; Buhrman,~R.~A.
	Current-Induced Switching of Domains in Magnetic Multilayer Devices.
	\emph{Science} \textbf{1999}, \emph{285}, 867--870\relax
	\mciteBstWouldAddEndPuncttrue
	\mciteSetBstMidEndSepPunct{\mcitedefaultmidpunct}
	{\mcitedefaultendpunct}{\mcitedefaultseppunct}\relax
	\EndOfBibitem
	\bibitem[Ralph and Stiles(2008)Ralph, and Stiles]{Ralph2008}
	Ralph,~D.~C.; Stiles,~M.~D. Spin transfer torques. \emph{J. Magn. Magn. Mater}
	\textbf{2008}, \emph{320}, 1190--1216\relax
	\mciteBstWouldAddEndPuncttrue
	\mciteSetBstMidEndSepPunct{\mcitedefaultmidpunct}
	{\mcitedefaultendpunct}{\mcitedefaultseppunct}\relax
	\EndOfBibitem
	\bibitem[Manchon \latin{et~al.}(2019)Manchon, \ifmmode~\check{Z}\else
	\v{Z}\fi{}elezn\'y, Miron, Jungwirth, Sinova, Thiaville, Garello, and
	Gambardella]{Manchon2019}
	Manchon,~A.; \ifmmode~\check{Z}\else \v{Z}\fi{}elezn\'y,~J.; Miron,~I.~M.;
	Jungwirth,~T.; Sinova,~J.; Thiaville,~A.; Garello,~K.; Gambardella,~P.
	Current-induced spin-orbit torques in ferromagnetic and antiferromagnetic
	systems. \emph{Rev. Mod. Phys.} \textbf{2019}, \emph{91}, 035004\relax
	\mciteBstWouldAddEndPuncttrue
	\mciteSetBstMidEndSepPunct{\mcitedefaultmidpunct}
	{\mcitedefaultendpunct}{\mcitedefaultseppunct}\relax
	\EndOfBibitem
	\bibitem[Rojas-S\'anchez \latin{et~al.}(2014)Rojas-S\'anchez, Reyren,
	Laczkowski, Savero, Attan\'e, Deranlot, Jamet, George, Vila, and
	Jaffr\`es]{Sanchez2013}
	Rojas-S\'anchez,~J.-C.; Reyren,~N.; Laczkowski,~P.; Savero,~W.;
	Attan\'e,~J.-P.; Deranlot,~C.; Jamet,~M.; George,~J.-M.; Vila,~L.;
	Jaffr\`es,~H. Spin Pumping and Inverse Spin {H}all Effect in Platinum: The
	Essential Role of Spin-Memory Loss at Metallic Interfaces. \emph{Phys. Rev.
		Lett.} \textbf{2014}, \emph{112}, 106602\relax
	\mciteBstWouldAddEndPuncttrue
	\mciteSetBstMidEndSepPunct{\mcitedefaultmidpunct}
	{\mcitedefaultendpunct}{\mcitedefaultseppunct}\relax
	\EndOfBibitem
	\bibitem[Zhang \latin{et~al.}(2015)Zhang, Han, Jiang, Yang, and
	Parkin]{Zhang2015}
	Zhang,~W.; Han,~W.; Jiang,~X.; Yang,~S.-H.; Parkin,~S. S.~P. Role of
	transparency of platinum-ferromagnet interfaces in determining the intrinsic
	magnitude of the spin {H}all effect. \emph{Nat. Phys.} \textbf{2015},
	\emph{11}, 496--502\relax
	\mciteBstWouldAddEndPuncttrue
	\mciteSetBstMidEndSepPunct{\mcitedefaultmidpunct}
	{\mcitedefaultendpunct}{\mcitedefaultseppunct}\relax
	\EndOfBibitem
	\bibitem[Dolui and Nikoli\ifmmode~\acute{c}\else \'{c}\fi{}(2017)Dolui, and
	Nikoli\ifmmode~\acute{c}\else \'{c}\fi{}]{Dolui2017}
	Dolui,~K.; Nikoli\ifmmode~\acute{c}\else \'{c}\fi{},~B.~K. Spin-memory loss due
	to spin-orbit coupling at ferromagnet/heavy-metal interfaces: Ab initio
	spin-density matrix approach. \emph{Phys. Rev. B} \textbf{2017}, \emph{96},
	220403\relax
	\mciteBstWouldAddEndPuncttrue
	\mciteSetBstMidEndSepPunct{\mcitedefaultmidpunct}
	{\mcitedefaultendpunct}{\mcitedefaultseppunct}\relax
	\EndOfBibitem
	\bibitem[Hirohata \latin{et~al.}(2016)Hirohata, Yamamoto, Murphy, and
	Vick]{Hirohata2016}
	Hirohata,~A.; Yamamoto,~Y.; Murphy,~B.~A.; Vick,~A.~J. Non-destructive imaging
	of buried electronic interfaces using a decelerated scanning electron beam.
	\emph{Nat. Commun.} \textbf{2016}, \emph{7}, 12701\relax
	\mciteBstWouldAddEndPuncttrue
	\mciteSetBstMidEndSepPunct{\mcitedefaultmidpunct}
	{\mcitedefaultendpunct}{\mcitedefaultseppunct}\relax
	\EndOfBibitem
	\bibitem[Jackson \latin{et~al.}(2018)Jackson, Sun, Kubota, Takanashi, and
	Hirohata]{Jackson2018}
	Jackson,~E.; Sun,~M.; Kubota,~T.; Takanashi,~K.; Hirohata,~A. Chemical and
	structural analysis on magnetic tunnel junctions using a decelerated scanning
	electron beam. \emph{Sci. Rep.} \textbf{2018}, \emph{8}, 1785\relax
	\mciteBstWouldAddEndPuncttrue
	\mciteSetBstMidEndSepPunct{\mcitedefaultmidpunct}
	{\mcitedefaultendpunct}{\mcitedefaultseppunct}\relax
	\EndOfBibitem
	\bibitem[Hirsch(1999)]{Hirsch1999}
	Hirsch,~J.~E. Spin {H}all Effect. \emph{Phys. Rev. Lett.} \textbf{1999},
	\emph{83}, 1834\relax
	\mciteBstWouldAddEndPuncttrue
	\mciteSetBstMidEndSepPunct{\mcitedefaultmidpunct}
	{\mcitedefaultendpunct}{\mcitedefaultseppunct}\relax
	\EndOfBibitem
	\bibitem[Sinova \latin{et~al.}(2015)Sinova, Valenzuela, Wunderlich, Back, and
	Jungwirth]{Sinova2015}
	Sinova,~J.; Valenzuela,~S.~O.; Wunderlich,~J.; Back,~C.~H.; Jungwirth,~T. Spin
	{H}all effects. \emph{Rev. Mod. Phys.} \textbf{2015}, \emph{87},
	1213--1260\relax
	\mciteBstWouldAddEndPuncttrue
	\mciteSetBstMidEndSepPunct{\mcitedefaultmidpunct}
	{\mcitedefaultendpunct}{\mcitedefaultseppunct}\relax
	\EndOfBibitem
	\bibitem[Berger \latin{et~al.}(2018)Berger, Edwards, Nembach, Karis, Weiler,
	and Silva]{Berger2018}
	Berger,~A.~J.; Edwards,~E. R.~J.; Nembach,~H.~T.; Karis,~O.; Weiler,~M.;
	Silva,~T.~J. Determination of the spin {H}all effect and the spin diffusion
	length of {P}t from self-consistent fitting of damping enhancement and
	inverse spin-orbit torque measurements. \emph{Phys. Rev. B} \textbf{2018},
	\emph{98}, 024402\relax
	\mciteBstWouldAddEndPuncttrue
	\mciteSetBstMidEndSepPunct{\mcitedefaultmidpunct}
	{\mcitedefaultendpunct}{\mcitedefaultseppunct}\relax
	\EndOfBibitem
	\bibitem[Brataas \latin{et~al.}(2006)Brataas, Bauer, and Kelly]{Brataas2006}
	Brataas,~A.; Bauer,~G. E.~W.; Kelly,~P.~J. Non-collinear magnetoelectronics.
	\emph{Phys. Rep.} \textbf{2006}, \emph{427}, 157--255\relax
	\mciteBstWouldAddEndPuncttrue
	\mciteSetBstMidEndSepPunct{\mcitedefaultmidpunct}
	{\mcitedefaultendpunct}{\mcitedefaultseppunct}\relax
	\EndOfBibitem
	\bibitem[Ando \latin{et~al.}(2011)Ando, Takahasi, Teda, Kajiwara, Nakayama,
	Yoshino, Harii, Fujikawa, Matsuo, Maekawa, and Saitoh]{Ando2011}
	Ando,~K.; Takahasi,~S.; Teda,~J.; Kajiwara,~Y.; Nakayama,~H.; Yoshino,~T.;
	Harii,~K.; Fujikawa,~Y.; Matsuo,~M.; Maekawa,~S.; Saitoh,~E. Inverse
	spin-{H}all effect induced by spin pumping in metallic system. \emph{J. Appl.
		Phys.} \textbf{2011}, \emph{109}, 103913\relax
	\mciteBstWouldAddEndPuncttrue
	\mciteSetBstMidEndSepPunct{\mcitedefaultmidpunct}
	{\mcitedefaultendpunct}{\mcitedefaultseppunct}\relax
	\EndOfBibitem
	\bibitem[Zhang \latin{et~al.}(2013)Zhang, Vlaminck, Pearson, Divan, Bader, and
	Hoffmann]{Zhang2013}
	Zhang,~W.; Vlaminck,~V.; Pearson,~J.~E.; Divan,~R.; Bader,~S.~D.; Hoffmann,~A.
	Determination of the {P}t spin diffusion length by spin-pumping and spin
	{H}all effect. \emph{Appl. Phys. Lett.} \textbf{2013}, \emph{103},
	242414\relax
	\mciteBstWouldAddEndPuncttrue
	\mciteSetBstMidEndSepPunct{\mcitedefaultmidpunct}
	{\mcitedefaultendpunct}{\mcitedefaultseppunct}\relax
	\EndOfBibitem
	\bibitem[Tao \latin{et~al.}(2018)Tao, Liu, B., Yu, Feng, Sun, You, Du, Chen,
	Zhang, Zhang, Yuan, Wu, , and Ding]{Tao2018}
	Tao,~X.; Liu,~Q.; B.,~M.; Yu,~R.; Feng,~Z.; Sun,~L.; You,~B.; Du,~J.; Chen,~K.;
	Zhang,~S.; Zhang,~L.; Yuan,~Z.; Wu,~D.; ; Ding,~H. Self-consistent
	determination of spin {H}all angle and spin diffusion length in {P}t and
	{P}d: The role of the interface spin loss. \emph{Sci. Adv.} \textbf{2018},
	\emph{4}, eaat1670\relax
	\mciteBstWouldAddEndPuncttrue
	\mciteSetBstMidEndSepPunct{\mcitedefaultmidpunct}
	{\mcitedefaultendpunct}{\mcitedefaultseppunct}\relax
	\EndOfBibitem
	\bibitem[Nakayama \latin{et~al.}(2013)Nakayama, Althammer, Chen, Uchida,
	Kajiwara, Kikuchi, Ohtani, Gepr\"ags, Opel, Takahashi, Gross, Bauer,
	Goennenwein, and Saitoh]{Nakayama2013}
	Nakayama,~H.; Althammer,~M.; Chen,~Y.-T.; Uchida,~K.; Kajiwara,~Y.;
	Kikuchi,~D.; Ohtani,~T.; Gepr\"ags,~S.; Opel,~M.; Takahashi,~S.; Gross,~R.;
	Bauer,~G. E.~W.; Goennenwein,~S. T.~B.; Saitoh,~E. Spin {H}all
	Magnetoresistance Induced by a Nonequilibrium Proximity Effect. \emph{Phys.
		Rev. Lett.} \textbf{2013}, \emph{110}, 206601\relax
	\mciteBstWouldAddEndPuncttrue
	\mciteSetBstMidEndSepPunct{\mcitedefaultmidpunct}
	{\mcitedefaultendpunct}{\mcitedefaultseppunct}\relax
	\EndOfBibitem
	\bibitem[Wang \latin{et~al.}(2014)Wang, Deorani, Qiu, Kwon, and Yang]{Wang2014}
	Wang,~Y.; Deorani,~P.; Qiu,~X.; Kwon,~J.~H.; Yang,~H. Determination of
	intrinsic spin {H}all angle in {P}t. \emph{Appl. Phys. Lett.} \textbf{2014},
	\emph{105}, 152412\relax
	\mciteBstWouldAddEndPuncttrue
	\mciteSetBstMidEndSepPunct{\mcitedefaultmidpunct}
	{\mcitedefaultendpunct}{\mcitedefaultseppunct}\relax
	\EndOfBibitem
	\bibitem[Nguyen \latin{et~al.}(2016)Nguyen, Ralph, and Buhrman]{Nguyen2016}
	Nguyen,~M.-H.; Ralph,~D.~C.; Buhrman,~R.~A. Spin Torque Study of the Spin
	{H}all Conductivity and Spin Diffusion Length in Platinum Thin Films with
	Varying Resistivity. \emph{Phys. Rev. Lett.} \textbf{2016}, \emph{116},
	126601\relax
	\mciteBstWouldAddEndPuncttrue
	\mciteSetBstMidEndSepPunct{\mcitedefaultmidpunct}
	{\mcitedefaultendpunct}{\mcitedefaultseppunct}\relax
	\EndOfBibitem
	\bibitem[Niimi \latin{et~al.}(2012)Niimi, Kawanishi, Wei, Deranlot, Yang,
	Chshiev, Valet, Fert, and Otani]{Niimi2012}
	Niimi,~Y.; Kawanishi,~Y.; Wei,~D.~H.; Deranlot,~C.; Yang,~H.~X.; Chshiev,~M.;
	Valet,~T.; Fert,~A.; Otani,~Y. Giant Spin {H}all Effect Induced by Skew
	Scattering from Bismuth Impurities inside Thin Film {C}u{B}i Alloys.
	\emph{Phys. Rev. Lett.} \textbf{2012}, \emph{109}, 156602\relax
	\mciteBstWouldAddEndPuncttrue
	\mciteSetBstMidEndSepPunct{\mcitedefaultmidpunct}
	{\mcitedefaultendpunct}{\mcitedefaultseppunct}\relax
	\EndOfBibitem
	\bibitem[Niimi \latin{et~al.}(2014)Niimi, Suzuki, Kawanishi, Omori, Valet,
	Fert, and Otani]{Niimi2014}
	Niimi,~Y.; Suzuki,~H.; Kawanishi,~Y.; Omori,~Y.; Valet,~T.; Fert,~A.; Otani,~Y.
	Extrinsic spin {H}all effects measured with lateral spin valve structures.
	\emph{Phys. Rev. B} \textbf{2014}, \emph{89}, 054401\relax
	\mciteBstWouldAddEndPuncttrue
	\mciteSetBstMidEndSepPunct{\mcitedefaultmidpunct}
	{\mcitedefaultendpunct}{\mcitedefaultseppunct}\relax
	\EndOfBibitem
	\bibitem[Sagasta \latin{et~al.}(2016)Sagasta, Omori, Isasa, Gradhand, Hueso,
	Niimi, Otani, and Casanova]{Sagasta2016}
	Sagasta,~E.; Omori,~Y.; Isasa,~M.; Gradhand,~M.; Hueso,~L.~E.; Niimi,~Y.;
	Otani,~Y.; Casanova,~F. Tuning the spin {H}all effect of {P}t from the
	moderately dirty to the superclean regime. \emph{Phys. Rev. B} \textbf{2016},
	\emph{94}, 060412\relax
	\mciteBstWouldAddEndPuncttrue
	\mciteSetBstMidEndSepPunct{\mcitedefaultmidpunct}
	{\mcitedefaultendpunct}{\mcitedefaultseppunct}\relax
	\EndOfBibitem
	\bibitem[Laczkowski \latin{et~al.}(2017)Laczkowski, Fu, Yang, Rojas-S\'anchez,
	Noel, Pham, Zahnd, Deranlot, Collin, Bouard, Warin, Maurel, Chshiev, Marty,
	Attan\'e, Fert, Jaffr\`es, Vila, and George]{Laczkowski2017}
	Laczkowski,~P. \latin{et~al.}  Large enhancement of the spin {H}all effect in
	{A}u by side-jump scattering on {T}a impurities. \emph{Phys. Rev. B}
	\textbf{2017}, \emph{96}, 140405\relax
	\mciteBstWouldAddEndPuncttrue
	\mciteSetBstMidEndSepPunct{\mcitedefaultmidpunct}
	{\mcitedefaultendpunct}{\mcitedefaultseppunct}\relax
	\EndOfBibitem
	\bibitem[Pham \latin{et~al.}(2016)Pham, Vila, Zahnd, Marty, Savero-Torres,
	Jamet, and Attan\'{e}]{Pham2016}
	Pham,~V.~T.; Vila,~L.; Zahnd,~G.; Marty,~A.; Savero-Torres,~W.; Jamet,~M.;
	Attan\'{e},~J.~P. Ferromagnetic/Nonmagnetic Nanostructures for the Electrical
	Measurement of the Spin {H}all Effect. \emph{Nano. Lett.} \textbf{2016},
	\emph{16}, 6755--6760\relax
	\mciteBstWouldAddEndPuncttrue
	\mciteSetBstMidEndSepPunct{\mcitedefaultmidpunct}
	{\mcitedefaultendpunct}{\mcitedefaultseppunct}\relax
	\EndOfBibitem
	\bibitem[Pfeiffer \latin{et~al.}(2018)Pfeiffer, Reeve, and
	Kl\"aui]{Pfeiffer2018}
	Pfeiffer,~A.; Reeve,~R.~M.; Kl\"aui,~M. Importance of spin current generation
	and detection by spin injection and the spin {H}all effect for lateral spin
	valve performance. \emph{J. Phys.: Condens. Matter} \textbf{2018}, \emph{30},
	465802\relax
	\mciteBstWouldAddEndPuncttrue
	\mciteSetBstMidEndSepPunct{\mcitedefaultmidpunct}
	{\mcitedefaultendpunct}{\mcitedefaultseppunct}\relax
	\EndOfBibitem
	\bibitem[Kimura \latin{et~al.}(2007)Kimura, Otani, Sato, Takahashi, and
	Maekawa]{Kimura2007}
	Kimura,~T.; Otani,~Y.; Sato,~T.; Takahashi,~S.; Maekawa,~S. Room-Temperature
	Reversible Spin {H}all Effect. \emph{Phys. Rev. Lett.} \textbf{2007},
	\emph{98}, 156601\relax
	\mciteBstWouldAddEndPuncttrue
	\mciteSetBstMidEndSepPunct{\mcitedefaultmidpunct}
	{\mcitedefaultendpunct}{\mcitedefaultseppunct}\relax
	\EndOfBibitem
	\bibitem[Jacquod \latin{et~al.}(2012)Jacquod, Whitney, Meair, and
	B\"uttiker]{Jacquod2012}
	Jacquod,~P.; Whitney,~R.~S.; Meair,~J.; B\"uttiker,~M. Onsager relations in
	coupled electric, thermoelectric, and spin transport: The tenfold way.
	\emph{Phys. Rev. B} \textbf{2012}, \emph{86}, 155118\relax
	\mciteBstWouldAddEndPuncttrue
	\mciteSetBstMidEndSepPunct{\mcitedefaultmidpunct}
	{\mcitedefaultendpunct}{\mcitedefaultseppunct}\relax
	\EndOfBibitem
	\bibitem[Kimura \latin{et~al.}(2008)Kimura, Sato, and Otani]{Kimura2008}
	Kimura,~T.; Sato,~T.; Otani,~Y. Temperature Evolution of Spin Relaxation in a
	$\mathrm{NiFe}/\mathrm{Cu}$ Lateral Spin Valve. \emph{Phys. Rev. Lett.}
	\textbf{2008}, \emph{100}, 066602\relax
	\mciteBstWouldAddEndPuncttrue
	\mciteSetBstMidEndSepPunct{\mcitedefaultmidpunct}
	{\mcitedefaultendpunct}{\mcitedefaultseppunct}\relax
	\EndOfBibitem
	\bibitem[Elliot(1954)]{Elliot1954}
	Elliot,~R.~J. Theory of the Effect of Spin-Orbit Coupling on Magnetic Resonance
	in Some Semiconductors. \emph{Phys. Rev.} \textbf{1954}, \emph{96},
	266--279\relax
	\mciteBstWouldAddEndPuncttrue
	\mciteSetBstMidEndSepPunct{\mcitedefaultmidpunct}
	{\mcitedefaultendpunct}{\mcitedefaultseppunct}\relax
	\EndOfBibitem
	\bibitem[Yafet(1963)]{Yafet1963}
	Yafet,~Y. G-factors and spin-lattice relaxation of conduction electrons.
	\emph{Sol. St. Phys.} \textbf{1963}, \emph{14}, 1--98\relax
	\mciteBstWouldAddEndPuncttrue
	\mciteSetBstMidEndSepPunct{\mcitedefaultmidpunct}
	{\mcitedefaultendpunct}{\mcitedefaultseppunct}\relax
	\EndOfBibitem
	\bibitem[Villamor \latin{et~al.}(2013)Villamor, Isasa, Hueso, and
	Casanova]{Villamor2013}
	Villamor,~E.; Isasa,~M.; Hueso,~L.~E.; Casanova,~F. Contribution of defects to
	the spin relaxation in copper nanowires. \emph{Phys. Rev. B} \textbf{2013},
	\emph{87}, 094417\relax
	\mciteBstWouldAddEndPuncttrue
	\mciteSetBstMidEndSepPunct{\mcitedefaultmidpunct}
	{\mcitedefaultendpunct}{\mcitedefaultseppunct}\relax
	\EndOfBibitem
	\bibitem[O'Brien \latin{et~al.}(2014)O'Brien, Erickson, Spivak, Ambaye,
	Goyette, Lauter, Crowel, and Leighton]{OBrien2014}
	O'Brien,~L.; Erickson,~M.~J.; Spivak,~D.; Ambaye,~H.; Goyette,~R.~J.;
	Lauter,~V.; Crowel,~P.~A.; Leighton,~C. Kondo physics in non-local metallic
	spin transport devices. \emph{Nat. Commun.} \textbf{2014}, \emph{5},
	3927\relax
	\mciteBstWouldAddEndPuncttrue
	\mciteSetBstMidEndSepPunct{\mcitedefaultmidpunct}
	{\mcitedefaultendpunct}{\mcitedefaultseppunct}\relax
	\EndOfBibitem
	\bibitem[Batley \latin{et~al.}(2015)Batley, Rosamond, Ali, Linfield, Burnell,
	and Hickey]{Batley2015}
	Batley,~J.~T.; Rosamond,~M.~C.; Ali,~M.; Linfield,~E.~H.; Burnell,~G.;
	Hickey,~B.~J. Spin relaxation through {K}ondo scattering in {C}u/{P}y lateral
	spin valves. \emph{Phys. Rev. B} \textbf{2015}, \emph{92}, 220420\relax
	\mciteBstWouldAddEndPuncttrue
	\mciteSetBstMidEndSepPunct{\mcitedefaultmidpunct}
	{\mcitedefaultendpunct}{\mcitedefaultseppunct}\relax
	\EndOfBibitem
	\bibitem[Kim \latin{et~al.}(2017)Kim, O'Brien, Crowell, Leighton, and
	Stiles]{Kim2017}
	Kim,~K.-W.; O'Brien,~L.; Crowell,~P.~A.; Leighton,~C.; Stiles,~M.~D. Theory of
	{K}ondo suppression of spin polarization in nonlocal spin valves. \emph{Phys.
		Rev. B} \textbf{2017}, \emph{95}, 104404\relax
	\mciteBstWouldAddEndPuncttrue
	\mciteSetBstMidEndSepPunct{\mcitedefaultmidpunct}
	{\mcitedefaultendpunct}{\mcitedefaultseppunct}\relax
	\EndOfBibitem
	\bibitem[Drouin \latin{et~al.}(2007)Drouin, Couture, Joly, Tastet, Aimez, and
	Gauvin]{Drouin2007}
	Drouin,~D.; Couture,~A.~R.; Joly,~D.; Tastet,~X.; Aimez,~V.; Gauvin,~R. CASINO
	{V}2.42-{A} Fast and easy-to-use Modeling Tool for Scanning Electron
	Microscopy and Microanalysis Users. \emph{Scanning} \textbf{2007}, \emph{29},
	92\relax
	\mciteBstWouldAddEndPuncttrue
	\mciteSetBstMidEndSepPunct{\mcitedefaultmidpunct}
	{\mcitedefaultendpunct}{\mcitedefaultseppunct}\relax
	\EndOfBibitem
	\bibitem[Kimura \latin{et~al.}(2006)Kimura, Otani, and Hamrle]{Kimura2006}
	Kimura,~T.; Otani,~Y.; Hamrle,~J. Enhancement of spin accumulation in a
	nonmagnetic layer by reducing junction size. \emph{Phys. Rev. B}
	\textbf{2006}, \emph{73}, 132405\relax
	\mciteBstWouldAddEndPuncttrue
	\mciteSetBstMidEndSepPunct{\mcitedefaultmidpunct}
	{\mcitedefaultendpunct}{\mcitedefaultseppunct}\relax
	\EndOfBibitem
	\bibitem[Lidig \latin{et~al.}(2019)Lidig, Cramer, Wei\ss{}hoff, Thomas,
	Kessler, Kl\"aui, and Jourdan]{Lidig2019}
	Lidig,~C.; Cramer,~J.; Wei\ss{}hoff,~L.; Thomas,~T.; Kessler,~T.; Kl\"aui,~M.;
	Jourdan,~M. Unidirectional Spin {H}all Magnetoresistance as a Tool for
	Probing the Interfacial Spin Polarization of
	$\mathrm{Co}_{2}\mathrm{Mn}\mathrm{Si}$. \emph{Phys. Rev. Applied}
	\textbf{2019}, \emph{11}, 044039\relax
	\mciteBstWouldAddEndPuncttrue
	\mciteSetBstMidEndSepPunct{\mcitedefaultmidpunct}
	{\mcitedefaultendpunct}{\mcitedefaultseppunct}\relax
	\EndOfBibitem
\end{mcitethebibliography}
\end{document}